\documentclass[aps,prl]{revtex4}
\usepackage{graphicx,color}
\usepackage{dcolumn}
\usepackage{bm}
\usepackage{amsmath,amsthm,amssymb}

\begin{document}

\title{
Charge current driven by spin dynamics
in disordered Rashba spin-orbit system
}

\author{Jun-ichiro Ohe}
\altaffiliation{%
Present address: Institute for Materials Research, Tohoku University,
Sendai 980-8577, Japan}
\email{johe@imr.tohoku.ac.jp}

\affiliation{%
I. Institut f\"{u}r Theoretische Physik, Universtit\"{a}t Hamburg,
Jungiusstrasse 9, 20355 Hamburg, Germany}

\author{Akihito Takeuchi}
\affiliation{%
Department of Physics, Tokyo Metropolitan University,
Hachioji, Tokyo 192-0397, Japan}

\author{Gen Tatara}
\affiliation{%
Department of Physics, Tokyo Metropolitan University,
Hachioji, Tokyo 192-0397, Japan \\
PRESTO, JST, 4-1-8 Honcho Kawaguchi, Saitama 332-0012, Japan
}


\date{\today}

\begin{abstract}
Pumping of charge current by spin dynamics in the presence of the Rashba spin-orbit interaction is theoretically studied. 
Considering disordered electron, the exchange coupling and spin-orbit interactions are treated perturbatively. 
It is found that dominant current induced by the spin dynamics is interpreted as a consequence of the conversion from spin current via the inverse spin Hall effect. 
We also found that the current has an additional component from a fictitious conservative field.
Results are applied to the case of moving domain wall.
\end{abstract}


\maketitle
\def\average#1{\langle {#1} \rangle}
\newcommand{\lt}{\left}
\newcommand{\rt}{\right}
\newcommand{\bv}{{\bm b}}
\newcommand{\DOS}{{\rho}}
\newcommand{\Ev}{{\bm E}}
\newcommand{\Eveff}{{\bm E}_{\rm eff}}
\newcommand{\Evind}{\Ev_{\rm ind}}
\newcommand{\ef}{\varepsilon_{\rm F}}
\newcommand{\evph}{{\bm e}_{\phi}}
\newcommand{\evth}{{\bm e}_{\theta}}
\newcommand{\gammaish}{{\gamma_{\rm ISH}}}
\newcommand{\Imag}{{\rm Im}}
\newcommand{\Jex}{{J_{\rm ex}}}
\newcommand{\js}{{j_{\rm s}}}
\newcommand{\jsv}{{\jv_{\rm s}}}
\newcommand{\jv}{{\bm j}}
\newcommand{\jphi}{{j^{\phi}}}
\newcommand{\jphimu}{{j^{\phi}_{\mu}}}
\newcommand{\jish}{{j^{\rm ISH}}}
\newcommand{\jishmu}{{j^{\rm ISH}_{\mu}}}
\newcommand{\gr}{g^{\rm r}}
\newcommand{\ga}{g^{\rm a}}
\newcommand{\kv}{{\bm k}}
\newcommand{\kf}{k_{\rm F}}
\newcommand{\mass}{{m}}
\newcommand{\nimp}{{n_{\rm i}}}
\newcommand{\phiind}{\phi_{\rm ind}}
\newcommand{\phis}{\bm{\phi}_{\rm s}}
\newcommand{\qv}{{\bm q}}
\newcommand{\Qv}{{\bm Q}}
\newcommand{\rhoind}{\rho_{\rm ind}}
\newcommand{\rashba}{{\alpha}}
\newcommand{\Real}{{\rm Re}}
\newcommand{\rv}{{\bm r}}
\newcommand{\sigmav}{\hat{\bm \sigma}}
\newcommand{\sigmaz}{{\sigma_0}}
\newcommand{\Sv}{{\bm S}}
\newcommand{\tr}{{\rm tr}}
\newcommand{\xv}{{\bm x}}
\newcommand{\xspin}{{\rm mid}}
\newcommand{\yspin}{{\rm hard}}
\newcommand{\zspin}{{\rm easy}}


Recent spintronics studies aim at manipulation both of
charge and spin degrees of freedom~\cite{Wolf}.
Central roles are played by the spin-orbit interaction and the exchange interaction between the conduction electrons and local spins~\cite{Berger96,Slonczewski96,Ohno,Rashba,Nitta}.
It has been shown that the exchange coupling is useful for electrical control of magnetization dynamics
via spin transfer torque~\cite{Berger96,Slonczewski96}.
It can also be used to pump spin current from the precession of the magnetization~\cite{Brataas00,Tserkovnyak02,Tserkovnyak05}.
The spin-orbit interaction has been recently found to induce a magnetism by the application of electric voltage (the spin Hall effect)~\cite{Murakami,Sinova,Kato}.

By combining the exchange and the spin-orbit interactions, various phenomena are expected, and the subject discussed in this paper is one of them; 
pumping of charge current by dynamical magnetization.
The idea is to convert the pumped spin current into a charge current by using of the spin-orbit interaction as proposed by Saitoh et al.~\cite{Saitoh06}.
This current due to the inverse spin Hall effect was indeed observed~\cite{Saitoh06,Valenzuela06,Kimura07} in metallic systems where the spin-orbit interaction is induced by Pt atom.

Theoretically, generation of the electric field or voltage due to the dynamical spin structure was discussed by Stern~\cite{Stern92}. 
The field is not a real electric field, but an effective one due to a spin Berry phase acting only on charge degrees of freedom with spin (like the electron).
The mechanism is similar to the Faraday's law, but a magnetic flux is replaced by a fictitious field from the spin Berry phase. 
The induced field is described by $\nabla\times \Eveff =-\dot{\bv}$, where $\bv$ is field of spin Berry phase, and current is divergenceless; $\nabla\cdot \jv=0$.
The theory was recently applied to a domain wall by Barnes and Maekawa~\cite{Barnes07}, and the effect of the spin relaxation was studied by Duine~\cite{Duine07}, 
where the relaxation was introduced by a phenomenological term ($\beta$-term).
All these studies have been done in the adiabatic limit, where the exchange coupling between the local spin and the conduction electrons is strong.

Charge current as a result of inverse spin Hall effect was theoretically studied by Zhang and Niu~\cite{Zhang04} and Hankiewicz et al.~\cite{Hankiewicz05} (the effect was called reciprocal spin Hall effect). 
The calculation was done  as a response to applied spin-dependent chemical potential, which would not be easy to control experimentally.
In contrast, our study tries to derive direct relation between physically accessible quantities, current and magnetization (local spin). 

Another type of a voltage generated at an interface of a ferromagnet-nonmagnet 
contact was predicted by Wang et al.~\cite{Wang06} in order to explain the experimental observation~\cite{Costache06}.
They have pointed out that a net charging is due to a back flow of spin current at the interface, and the spin accumulation at the interface is essential.

In this letter, we theoretically predict another mechanism of a spin-induced charge battery realized in disordered conductors.
We consider the perturbative regime of the exchange coupling,
that is in the opposite limit of adiabatic cases~\cite{Stern92,Barnes07,Duine07}.
Proposed mechanism does not rely on the interface, and the pumped current arises simply when the exchange interaction and the spin-orbit interaction exist simultaneously.

We consider the two-dimensional electron gas system (2DEGs)
with the Rashba spin-orbit interaction that originates from
the lack of the inversion symmetry~\cite{Rashba,Nitta,Yanase07}.
The conduction electrons couples to the local magnetic moment
via the exchange coupling.
Such a system can be achieved by the 2DEGs
attached to the ferromagnetic contact~\cite{Datta,Hanbicki,Matsuyama},
or the 2DEGs in magnetic semiconductors (e.g., CdTe/CdMnTe)~\cite{Scholl,Takano}.

Let us consider the current representation by the simple argument.
The current $j_\mu$, proportional to the average 
$\tr \average{k_\mu}$ of the electron wave vector with respect to electron states ($\tr$ is trace over spin indices), vanishes if the system is spatially symmetric.
The Rashba interaction, proportional to $(\kv\times\sigmav)_z$,
breaks the spatial symmetry,
but charge current does not arise since 
$\tr \average{k_\mu (\kv\times\sigmav)_z}=0$ (only spin current arises~\cite{Brataas00}).
Charge current appears when we introduce the exchange coupling, proportional to $\Sv\cdot\sigmav$ where $\Sv$ is a local spin. 
We would have the current
$j_\mu\propto \tr \average{k_\mu (\kv\times\sigmav)_z\Sv\cdot\sigmav}
\propto \epsilon_{\mu\nu z}S^\nu$ to the first order exchange interaction,
and $j_\mu\propto  \epsilon_{\mu\nu z}(\Sv\times\Sv')^\nu$ 
at the second oder interaction with different spins, 
$\Sv$ and $\Sv'$.

In order to obtain the precise expression of the current, we perform
analytical calculations by using a diagrammatic technique.
Although the argument in the previous paragraph gives the qualitative idea of
pumping charge current, it turns out that the linear term in $\Sv$ vanishes
identically in the Rashba case, and the second order term needs a dynamical part as $\Sv\times\dot{\Sv}$.
We will demonstrate that the pumped current has two components.
One describes the inverse spin Hall effect~\cite{Saitoh06},
and the other is a conservative current which written as a divergence of
a scalar potential.
We will show that the current due to the inverse spin Hall effect is
dominant in various cases, and the conservative current is
relatively small.
However, the conservative one is also important from the view of the
fundamental physics, because it provides
the fictitious scalar potential which acts only on the
particle having both charge and spin.

We consider a disordered 2DEGs with the Rashba spin-orbit interaction.
The 2DEGs also interact with the local spin, $\Sv_{\xv}(t)$, via the exchange coupling.
The local spin is treated as classical, and slowly varying in space and time.
The system is represented by a Hamiltonian  
$H=H_0+H_{\rm{ex}}(t)+H_{\rm{so}}+H_{\rm imp}$, 
where
$H_0\equiv\sum_\kv \varepsilon_\kv c_\kv^{\dagger}c_\kv$
describes free electrons 
with 
$\varepsilon_{\kv}=\hbar^2\kv^2/2 \mass$
($\mass$ being the effective mass).
The exchange and the Rashba spin-orbit interactions are given by
\begin{eqnarray}\label{H}
H_{\rm ex}(t)&=&-J_{\rm ex}\sum_{\kv,\qv,\Omega} 
\Sv_\qv(\Omega)e^{\mathrm{i}\Omega t}
 c^{\dagger}_{\kv+\qv}
\sigmav c_\kv, \nonumber \\
H_{\rm so}&=& -\rashba \sum_{\kv}
\epsilon_{\mu\nu z} k_\mu (c_{\kv}^{\dagger} \hat{\sigma}^\nu c_\kv),
\end{eqnarray}
where 
$\Jex$ is the exchange coupling constant, 
$\Sv_\qv(\Omega)$ denotes the Fourier transform of the
local spin structure and
$\rashba$ is the strength of the Rashba spin-orbit interaction.
Spin-independent disorder is represented by 
$H_{\rm imp}=\sum_{i=1}^{\nimp}\sum_{\kv,{\kv'}}\frac{u}{V}
e^{\mathrm{i}({\kv}-{\kv'})\cdot \rv_i}c_{\kv'}^{\dagger}c_\kv$, 
which gives rise an elastic electron lifetime 
$\tau=(2\pi \DOS \nimp u^2/V)^{-1}$, where $\DOS$ is the density of states, 
$\nimp$ is the number of the impurities,
$u$ is the strength of the impurity scattering
and $V$ is the volume of the system.

The charge current density of this system is given by 
\begin{eqnarray}\label{jdef}
j_{\mu}(\xv,t)
&=& \frac{\mathrm{i}e}{V} \sum_{{\kv},{\kv'}}
e^{\mathrm{i}({\kv}-{\kv'})\cdot\xv}
\tr \lt[ \lt( \frac{\lt(\kv+\kv'\rt)_{\mu}}{2\mass} 
-\alpha\epsilon_{\mu\nu z} \hat{\sigma}^{\nu} \rt) 
G_{{\kv},{\kv'}}^{<}(t,t) \rt].
\end{eqnarray}
$G_{\kv,\kv'}^{<}(t,t')$ is a lesser Green function
which is
a $2\times2$ matrix in spin space with components  $G_{\kv\sigma,\kv'\sigma'}^{<}(t,t')=
\mathrm{i}\average{c_{\kv'\sigma'}^{\dagger}(t')c_{\kv\sigma}(t)}$ 
($\sigma,\sigma'=\pm$),
where $\average{\cdots}$ is the expectation value estimated by the total Hamiltonian $H$.

\begin{figure}[bt]
\begin{center}
\includegraphics[scale=0.2]{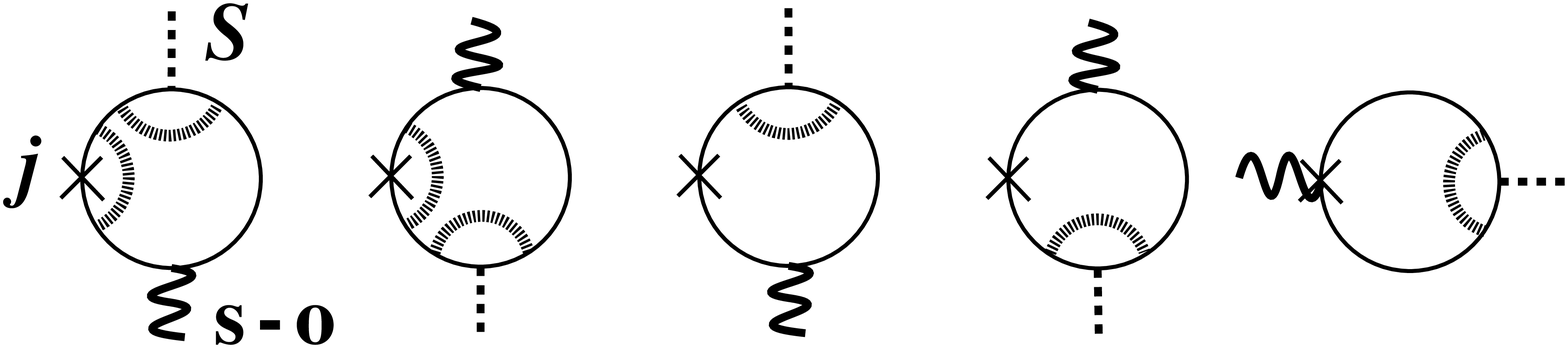}
\caption{Diagrammatic representation of currents at the first order in $\Jex$ and the lowest order in $\Omega$, which turns out to vanish. 
Dotted lines and wavy lines denote local spins $S$ and the Rashba interaction, respectively, and thick line represents the diffusion ladder, $D_q$. 
\label{FIGdiag1}
}
\end{center}
\end{figure}

\begin{figure}[bt]
\begin{center}
\includegraphics[scale=0.2]{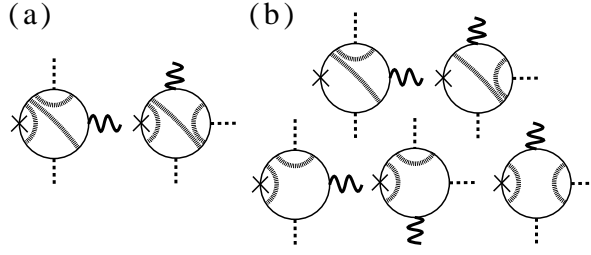}
\caption{Leading contribution from the second order in $\Jex$. 
Contributions from (a) and (b) are denoted by $\jphimu$ and $\jishmu$
in Eq.~(\ref{jres}), respectively.
Contributions from other diagrams cancel out or are smaller by $\mathcal{O}(1/\ef \tau)$.
}
\label{FIGdiag2}
\end{center}
\end{figure}

We calculate the current by treating both Rashba (to the first order) and exchange interactions perturbatively, which is valid if 
$\Jex\tau\ll1$ and $\alpha\kf\tau \ll1$~\cite{Inoue04}.
Successive impurity scatterings are denoted by ladder approximation, resulting in a diffusion propagator at small momentum transfer ($q$), $D_q \equiv (Dq^2 \tau)^{-1}$, where $D \equiv \kf^2\tau/2m^2$ is a diffusion constant. 
Dominant contributions are from diagrams that include a maximal number of diffusion propagators.
Contributions from the first order in $\Jex$ are shown in Fig.~\ref{FIGdiag1},
that turn out to vanish identically.
The leading contribution coming from the second order in $\Jex$ are shown in Fig.~\ref{FIGdiag2}.

After straightforward calculations, the current in the slowly varying limit ($\Omega\tau \ll1$) is obtained as
\begin{eqnarray}
j_\mu(\xv,t) &=& \frac{4e\rashba \Jex^2}{\mathrm{i}\pi mV} \epsilon_{\nu\eta z} (\tfrac{\nimp u^2}{V}) \sum_{\qv,\Qv} e^{-\mathrm{i}\Qv\cdot\xv} D_q
(\Sv_{\Qv-\qv} \times \dot{\Sv}_{\qv})_\eta \nonumber\\
&& \times 
\lt[ (\tfrac{\nimp u^2}{V}) D_q D_Q A^\mu_{\Qv} B^\nu_{\qv} C_{\qv\Qv}
+D_q E^\mu_{\qv\Qv} B^\nu_{\qv}
-D_Q A^\mu_{\Qv} F^\nu_{\qv\Qv} \rt],
\label{j1}
\end{eqnarray}
where 
$A^\mu_{\Qv} \equiv \sum_{\kv}k_\mu \gr_{\kv-\frac{\Qv}{2}}\ga_{\kv+\frac{\Qv}{2}}$, 
$B^\nu_{\qv} \equiv \sum_{\kv}k_\nu (\gr_{\kv})^2\ga_{\kv+\qv}$, 
$C_{\qv\Qv} \equiv \Real\sum_{\kv}\gr_{\kv}\ga_{\kv+\qv}\ga_{\kv+\Qv}$,
$E^\mu_{\qv\Qv} \equiv \mathrm{i}\Imag\sum_{\kv}\lt(\kv+\frac{\Qv}{2}\rt)_\mu\gr_{\kv}\ga_{\kv+\qv}\ga_{\kv+\Qv}$ and
$F^\nu_{\qv\Qv} \equiv \Real\sum_{\kv}(\kv+\qv)_\nu \gr_{\kv}(\ga_{\kv+\qv})^2\ga_{\kv+\Qv}$
($\gr_{\kv}=(\ga_{\kv})^*=(\ef-\varepsilon_{\kv}+\frac{\mathrm{i}}{2\tau})^{-1}$ ($\ef$ being the Fermi energy)).
The second and third terms in Eq.~(\ref{j1}) (proportional to $E^\mu_{\qv\Qv}$ and $F^\nu_{\qv\Qv}$, respectively) are the second leading term with a long range limit ($q, Q \rightarrow0$).
It is, however, physically the most essential term as we see below. 
For a spatially smooth structure of spins, i.e., $q, Q \ll \ell^{-1}$ ($\ell$ is the electron mean free path),
 we can approximate 
$A^\mu_{\Qv} \sim 2\pi \mathrm{i} \DOS \tau m D Q_\mu$, 
$B^\nu_{\qv} \sim 4\pi \DOS \tau^3 \ef q_\nu$, 
$C_{\qv\Qv} \sim -\DOS/2\ef^2$, 
$E^\mu_{\qv\Qv} \sim -\pi \mathrm{i} \DOS \tau^2 q_\mu$ and
$F^\nu_{\qv\Qv} \sim \pi \DOS \tau^3 Q_\nu$.
Then, the current is obtained as~\cite{jnote} 
$j_\mu(\xv,t) =\jishmu(\xv,t) +\jphimu(\xv,t) $, where 
\begin{eqnarray}
\jishmu(\xv,t) &=& 
- \frac{3 em \rashba \Jex^2 \tau^2}{\pi}
 \epsilon_{\mu\eta z} \int \frac{d^2x_1}{a^2} D_{\xv-\xv_1}
({\Sv}_{\xv} \times \dot{\Sv}_{\xv_1})_\eta,
\nonumber\\
\jphimu(\xv,t) &=& 
\frac{2 e \rashba \Jex^2 \tau^3}{\pi^2}
\epsilon_{\nu \eta z} \frac{\partial}{\partial x_\mu}
\int \frac{d^2x_1}{a^2} \int \frac{d^2x_2}{a^2}  D_{\xv-\xv_1}
 \frac{\partial D^{(2)}_{\xv_1-\xv_2}}{\partial x_{1\nu}}
({\Sv}_{\xv_1} \times \dot{\Sv}_{\xv_2})_\eta.
\label{jres}
\end{eqnarray}
Here, $D_\xv\equiv \frac{a^2}{V}\sum_{\qv}e^{-\mathrm{i}\qv\cdot\xv}D_\qv$,
$D^{(2)}_\xv\equiv \frac{a^2}{V} \sum_{\qv}e^{-\mathrm{i}\qv\cdot\xv}(D_\qv)^2$ and $a$ is the lattice constant.
(Note that singular behavior at $q\rightarrow0$ is cut off at
$q\sim L^{-1}$, where $L$ is a system size.)

The first term in Eq.~(\ref{jres}), $\jish$, describes a current whose direction is correlated with the magnetization direction, perpendicular to $\Sv\times\dot{\Sv}$,
and represents inverse spin Hall effect~\cite{Saitoh06}.
In order to make clear the physical meaning of this current,
we compare with the spin currents pumped by magnetization.
In the absence of spin-orbit interaction, we obtain the pumped spin current
at the lowest order in $\Jex$ as
\begin{eqnarray}
\jsv_{\mu}(\xv,t) &=&
\frac{\Jex \ef \tau^2}{2\pi}
\frac{\partial}{\partial x_\mu}
\int \frac{d^2x_1}{a^2} D_{\xv-\xv_1}
\lt[ \dot{\Sv}_{\xv_1} 
-2 \Jex \tau \int \frac{d^2x_2}{a^2} D_{\xv_1-\xv_2}
(\Sv_{\xv_1} \times \dot{\Sv}_{\xv_2}) \rt].
\label{jsres}
\end{eqnarray}
The polarization of the spin current is in both directions, 
$\propto \average{\dot{\Sv}}$ and $\average{\Sv\times \dot{\Sv}}$,
where $\average{\cdots}$ denotes average over diffusive electron motion.
This spin current is a gradient of a certain spin potential, $\jsv_\mu = -\nabla_\mu \phis$.
The result of Eq.~(\ref{jsres}) is consistent with the observation by
Tserkovnyak et al.~\cite{Tserkovnyak02}, where the spin current appears
 at the interface between ferromagnet and normal metals
associated with the phenomenological parameter of spin-mixing conductance.
(We note that there is also an equilibrium component of spin current~\cite{TK03}. This component,
$\jsv^{\rm (eq)}_\mu(\xv)= \frac{\Jex^2}{24\pi^2\ef^2\tau} (\nabla_\mu\Sv_{\xv}) \times \Sv_{\xv}$, is free from diffusion poles.
Hence, it is local and therefore small compared with dynamical contributions.)
The meaning of pumped spin current is understood by taking a divergence:
\begin{eqnarray}
\nabla\cdot\jsv(\xv,t) &=&
-\frac{m \Jex}{2\pi}
\lt[ \dot{\Sv}_{\xv}
-2\Jex\tau \int \frac{d^2 x_1}{a^2} D_{\xv-\xv_1}(\Sv_{\xv} \times \dot{\Sv}_{\xv_1}) \rt]. \label{jsconsv}
\end{eqnarray}
Comparing the second term ($\nabla\cdot{\js}^{(2)\nu}$) to $\jish$, we see that
$\jishmu = \gammaish \epsilon_{\mu\nu z} (\nabla \cdot {\js}^{(2)\nu})$,
where $\gammaish = -3 e \rashba \tau$.
This expression is the Rashba-version of inverse spin Hall effect, 
$\jv\propto \jv_s \times \sigmav$,
proposed in Ref.~\cite{Saitoh06}.
(Note that the spin current considered in Ref.~\cite{Saitoh06} is the one flowing through the interface that enables the spin current to enter without divergence.)
Eq.~(\ref{jsconsv}) represents a conservation law of spin, and correctly describes a fact that the Gilbert-type damping ($\average{\Sv\times \dot{\Sv}}$) results in a flow of spin current or $\dot{\Sv}$.

\begin{figure}[bt]
\begin{center}
\includegraphics[scale=0.3]{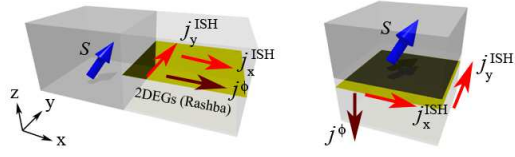}
\caption{(Color online) Two typical geometries with ferromagnets attached to two-dimensional electron system.
Left: Datta-Das geometry~\protect\cite{Datta} and Right: perpendicular geometry like in Ref.~\protect\cite{Saitoh06}.
In both cases, inverse spin Hall current is given by
$\jishmu\propto \epsilon_{\mu\nu z} (\Sv\times\dot{\Sv})_{\nu}$ ($\mu=x,y$).
\label{FIGsystem}
}
\end{center}
\end{figure}

In contrast, the second term in Eq.~(\ref{jres}), $\jphi$, is a gradient of a scalar quantity.
It can be interpreted as a current arising from a potential or a conserved force.
The fictitious electric field, defined by 
$\Evind= \jv^{\phi}/\sigmaz$, where $\sigmaz=e^2 n\tau/m$ is Boltzmann conductivity, is written as $\Evind=-\nabla \phiind$.
The scalar potential is obtained as
\begin{eqnarray}
\phiind(\xv,t) &=& -\frac{2\rashba\Jex^2\tau^2}{\pi e \ef}
 \epsilon_{\nu\eta z} 
\int \frac{d^2x_1}{a^2} \int \frac{d^2x_2}{a^2}  D_{\xv-\xv_1}
\frac{\partial D^{(2)}_{\xv_1-\xv_2}}{\partial x_{1\nu}}
({\Sv}_{\xv_1} \times \dot{\Sv}_{\xv_2})_\eta.
\end{eqnarray}
(Note that these scalar potential and field are fictitious ones, acting only on charge having spin degrees of freedom.)
The current $\jphi$ is in the direction where
magnetization changes.
It contributes to the
perpendicular current in the Datta-Das spin transistor geometry~\cite{Datta}.
However, it is not in-plane current in the
layer geometry~\cite{Saitoh06} as shown in
Fig.~\ref{FIGsystem}. 
It does not contribute to the case of moving domain walls
as we will show below.

\begin{figure}[bt]
\begin{center}
\includegraphics[scale=0.3]{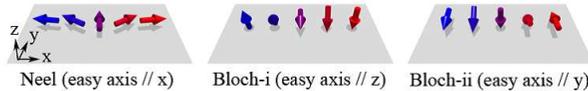}
\caption{(Color online) Three domain walls in the $xy$-plane,
Neel, Bloch-i and Bloch-ii.
\label{FIGdws}
}
\end{center}
\end{figure}

Let us apply our results to a case of a moving domain wall (as would be realized by using magnetic semiconductors~\cite{Scholl,Takano}).
We define polar angles $(\theta,\phi)$ with respect to the easy and hard axis of spin as $\cos\theta=S_{\zspin}/S, \tan\phi=S_{\yspin}/S_{\xspin}$, where $S_\zspin$, $S_\yspin$ and $S_\xspin$ denote spin components in easy-, hard- and medium-anisotropy directions.
Rigid and one-dimensional (in the $x$-direction) domain wall solution is represented by two dynamical variables, $X(t)$ and $\phi(t)$~\cite{Slonczewski72}, as
$\cos\theta(x,t)=\tanh\frac{x-X(t)}{\lambda}$ and
$\sin\theta(x,t)=[\cosh\frac{x-X(t)}{\lambda}]^{-1}$, where $\lambda$ is wall thickness.
By assuming a dirty case $\lambda \gg \ell$ ($\ell$ being the mean free path), and noting that $\Sv_{\xv}\times\dot{\Sv}_{\xv'}$ vanishes if $|\xv-\xv'|\gg \lambda$, we can approximate $\Sv_{\xv}\times\dot{\Sv}_{\xv'}$ by local value as $ \Sv_{\xv}\times\dot{\Sv}_{\xv}
\sim \sin\theta(x,t)(\dot{\phi} \evth -\frac{\dot{X}}{\lambda}\evph)$.
Here, $\evth=(\cos\theta\cos\phi, \cos\theta\sin\phi, -\sin\theta)$ and $\evph=(-\sin\phi,\cos\phi,0)$, indicating
that the damping ($\Sv\times \dot{\Sv}$) on the translational motion ($\dot{X}$) is in the $\phi$-direction, while it is within the wall plane when $\phi$ varies.
The pumped inverse spin Hall current,
$\jishmu\sim -j_0
 \epsilon_{\mu\eta z} 
\average{\Sv \times \dot{\Sv}}_\eta$,
where $j_0\equiv \frac{3}{\pi} em\rashba\Jex^2 \tau^2 D_0$ ($D_0\equiv \frac{L}{a^2} \int_{\ell}^{\lambda}dx D_x$)
depends much on the wall geometry.
We consider three types of the domain wall as shown in Fig.~\ref{FIGdws}, a Neel wall (N) and two Bloch walls ((B-i) and (B-ii)).
By estimating $\average{\Sv \times \dot{\Sv}}$ inside the wall (by using $\average{\cos\theta}=\average{\tanh\frac{x}{\lambda}}=0$ etc.), we obtain 
\begin{eqnarray}
{j^{\rm ISH}_x }^{\rm (N)} \sim - j_0 \frac{\dot{X}}{L} \sin\phi, \;\;\; 
{j^{\rm ISH}_x }^{\rm (B-i)} \sim j_0 \frac{\dot{X}}{L} \cos\phi, \label{jn}
\end{eqnarray}
which are driven by translational motion, and
\begin{eqnarray}
{j^{\rm ISH}_x }^{\rm (B-ii)} \sim j_0 \frac{\lambda}{L} \dot{\phi}, \label{jb}
\end{eqnarray}
which is driven by tilt of the wall.
In the case of the Neel and the out-of-plane Bloch wall,  
the current is a constant (if wall velocity is constant) at small speed and shows an oscillation ($\sin \phi$ or $\cos \phi$) when the domain wall is above Walker's break down. 
In contrast, no current is induced  for the in-plane Bloch wall
at small velocity and finite but steady current arises above breakdown (as long as $\dot\phi$ is more or less constant). 
The gradient part of the current, $\jphi$, on the other hand, is averaged out for any type of the domain wall. 

Let us briefly see the magnitude of current (Eqs.~(\ref{jn})(\ref{jb})). 
We use
$\ef=10$meV, $\Jex=10$meV, $a=10$nm and 
$\rashba=0.3\times 10^{-11}$eVm~\cite{Nitta}.
Using $D_0\sim (L^3/a^2\ell^2\lambda)\ln(L/\lambda)$ 
(our diffusive result depends much on sample size $L$), the current is estimated as 
$j\sim 4\times 10^{-14}$[C/m]$\times\frac{L^2\lambda^2}{a^4}
\ln\frac{L}{\lambda}\times\Omega$[Hz].
If we choose $\Omega=100$MHz, $L=1\mu$m and $\lambda\sim10a$, we obtain $j\sim 10$[A/m],
i.e., current is $I\equiv jL\sim 10\mu$A, which would be detectable experimentally.

We have considered a case of an uniform Rashba interaction. 
If we allow $\rashba$ to be position dependent, $\rashba(\xv)$,
we have other contributions. 
For instance, charge current linear in $\Sv$ arises proportional to 
$\average{(\nabla \rashba) \dot{\Sv}}$.
Thus, various currents are expected in the case of finite-size Rashba system (finite size is always the case in experiments).

In conclusion, we have theoretically shown that 
charge current is pumped by magnetization dynamics in the presence of
the Rashba spin-orbit interaction. 
The dominant part was found to be due to the inverse spin Hall effect, i.e., conversion of spin current into charge current by spin-orbit interaction.
In addition to the inverse spin Hall current, we found 
a conservative current flowing basically along the
gradient of the magnetization damping.
This current is rotation free, and should be distinguished from the
inverse spin Hall current and from the divergenceless current predicted
by Stern and others~\cite{Stern92,Barnes07,Duine07}.
It would be extremely interesting if one could
experimentally determine the type of the pumped current.

\begin{acknowledgments}
The authors are grateful to E. Saitoh, B. Kramer, R. Raimondi, M. Yamamoto,
S. Kettemann, J. Shibata, H. Kohno, S. Murakami,  and T. Ohtsuki
for valuable discussions.
This work has been supported by the
Deutsche Forschungsgemeinschaft via SFBs 508 and 668 of the
Universit\"at Hamburg.

{\it Noted added in proof.}
---After finishing the manuscript, we found optically induced inverse
spin Hall effect was observed in GaAs\cite{Zhao}.

\end{acknowledgments}


\begin{thebibliography}{99}
\bibitem{jnote}
Correctly speaking, even without spin-orbit interaction, small charge current arises at the second order of diffusion pole.
This current, proportional to $\average{\Sv_{\xv}\cdot\dot{\Sv}_{\xv'}}$, is of non-magnetic origin, i.e., due to time-dependent potential, and is of minor importance, besides being very small when $\Sv$ is slowly varying in space, $\Sv\cdot\dot{\Sv}\sim 0$.
\end{thebibliography}

\end{document}